\documentclass[12pt]{article}
\usepackage{graphicx}
\usepackage{natbib}

\def\fracd#1#2{{\displaystyle\frac{#1}{#2}}}
\def\sumd{\displaystyle\sum}

\begin{document}

\title{On computation of a common mean}
\author{Zinovy Malkin}
\date{Pulkovo Observatory, Pulkovskoe Ch. 65, St. Petersburg 196140, Russia \\ e-mail: malkin@gao.spb.ru}
\maketitle

\begin{abstract}
Combining several independent measurements of the same physical quantity is one of the most important tasks in metrology.
Small samples, biased input estimates, not always adequate reported uncertainties, and unknown error distribution make a rigorous solution very
difficult, if not impossible.
For this reason, many methods to compute a common mean and its uncertainty were proposed, each with own advantages and shortcomings.
Most of them are variants of the weighted average (WA) approach with different strategies to compute WA and its standard deviation.
Median estimate became also increasingly popular during recent years.
In this paper, these two methods in most widely used modifications are compared using simulated and real data.
To overcome some problems of known approaches to compute the WA uncertainty, a new combined estimate has been proposed.
It has been shown that the proposed method can help to obtain more robust and realistic estimate suitable for both consistent and discrepant measurements.
\end{abstract}

\section{Introduction}

Computation of a common mean (CM) of several independent measurements of a physical quantity is a common procedure in scientific analysis.
Typical task include combining results obtained by different analysts or results based on different measurement methods or results of several series
of measurements, etc.
One of the most important applications required computation of a CM is derivation of the best estimate of physical constants.
Both accurate CM value and its realistic uncertainty are equally important in such computations.

Input information consists of measured values $x_i$, $i=1,\ldots,n$ with associated uncertainties $s_i$, usually standard deviations (STD),
and correlations between $x_i$ are not available.
From statistical point of view, it is a classical case of direct measurements of unequal precision.
Due to lack of needed information we have to treat them as uncorrelated.
Combination procedure aimed to get an estimate $\bar{x}$ of the CM $x$ with associated uncertainty $\sigma$, which adequately reflects the scatter
and uncertainties of input measurements.

There is no unambiguous solution of this problem.
It is well known the the classical weighted average (WA) is unbiased CM estimate with minimum variance, provided $x_i$ are independent unbiased
estimators, and $s_i^2$ are true variances.
A solution of the problem is not straightforward if true variances are unknown.
Numerous papers are devoted to computation of a CM and its uncertainty; see e.g. \citet{Graybill1959,Sinha1985,Witkovsky2001, Zhang2006} and references therein.
Unfortunately, results are obtained under rather strong assumptions: the $x_i$ are unbiased estimates of $\bar{x}$, $s_i$ are normally distributed,
and number of measurements $n_i$ formed each $x_i$ is known.
Evidently, these assumptions are hardly met in scientific data analysis.
The situation becomes even more complicated because small samples are are the rule rather than the exception in such tasks,
which makes it difficult to efficiently apply most of statistical methods generally used.

For these reasons many alternative approaches, not always strongly justified, are mostly used, see e.g. \citet{MacMahon2004}, \citet{Chen2011},
and Dataplot software documentation\footnote{http://www.itl.nist.gov/div898/software/dataplot/}.
We do not aim this paper at investigation of all the existing methods.
Such a task seems to be impractical because some methods cannot be used in our case because we lack needed information, some are developed for specific
applications, and some approaches may be too complicated for routine use, e.g. bootstrapping \citep{Helene2007} or total median 
\citep{Figueiredo2004,Cox2004}.

We consider two basic methods most commonly used: WA and median.
As to the former, several approaches to compute its uncertainty (STD) are proposed in literature.
This paper is devoted to comparison of this methods based on simulated and real data.
Besides, a new combined approach is proposed to compute the WA uncertainty.
This approach was proposed for the first time in \citet{Malkin2001k} and has shown to be useful in practical applications such as computation of Earth
orientation parameter (EOP) combined solution \citep{Malkin2001m}, radio source position catalogues combination \citep{Sokolova2007},
analysis of radio source position time series \citep{Malkin2008g}, and modeling the galactic aberration \citep{Malkin2011fe}.
Special attention is given to analysis of small statistical samples.

\section{Computation of CM}

\subsection{Basic WA estimators}

The WA estimator is most widely used in various scientific and practical applications.
Let we have $n$ values $x_i$ with associated standard deviations $s_i$, $i=1 \ldots n$.
Then we can compute the following statistics (e.g. \citet{Brandt1999}; \citet{Bevington2003})
\begin{equation}
p_i = \fracd{1}{s_i^2} \ , \quad p = \sumd_{i=1}^n {p_i} \ , \quad \bar{x}_w = \fracd{\sumd_{i=1}^n  p_i x_i}{p} \ .
\label{eq:basic}
\end{equation}

In result, we have the classical WA estimate $\bar{x}_w$ with weights inversely proportional to variances of input measurements $s_i^2$.
We also can compute the following statistics also used as a measure of goodness of fit
\begin{equation}
H = \sumd_{i=1}^n {p_i (x_i - \bar{x}_w)^2} = \sumd_{i=1}^n {\left[\fracd{(x_i - \bar{x}_w)}{s_i}\right]^2} \ ,
\label{eq:h}
\end{equation}
where $H$ has a $\chi^2$ distribution with $n-1$ degree of freedom ($dof$) if $s_i^2$ are theoretical variances.
In practice, $s_i^2$ are, as a rule, sample variances, but this fact is usually ignored.
The $H$ statistics is an indicator of consistency of input measurements.
If the measurements are consistent, the value
\begin{equation}
\fracd{\chi^2}{dof} = \fracd{H}{n-1} \\
\label{eq:chi2}
\end{equation}
is close to unity.
Otherwise one can assume that the input measurements $x_i$ have systematic errors or $s_i$ are underestimated.

The question is how to estimate the standard error $\sigma$ of the mean?
Two main approaches can be applied to compute $\sigma$.
The classical WA estimate is
\begin{equation}
\sigma_1 = \fracd{1}{\sqrt{p}} \ .
\label{eq:sigma_1}
\end{equation}

Least squares approach leads to alternative estimate of the WA uncertainty.
The solution of the least square problem $x_i=\bar{x}+\varepsilon_i$ with weights $p_i$ gives the same WA estimate $\bar{x}$, but
another estimate of its uncertainty:
\begin{equation}
\sigma_2 = \sqrt{\fracd{\sumd_{i=1}^n {p_i (x_i - \bar{x}_w)^2}}{p\,(n-1)}} \ .
\label{eq:sigma_2}
\end{equation}

This estimate of the WA uncertainty is, in fact, $\sigma_1$ estimate scaled in such a way to make $\chi^2/dof$ close to unity:
\begin{equation}
\sigma_2 = \sigma_1\,\sqrt{\fracd{H}{n-1}} \ ,
\label{eq:sigma_2H}
\end{equation}
which is equivalent to scaling of input $s_i$ by the factor $\sqrt{H/(n-1)}$ as recommended by \citet{Rosenfeld1967} and \citet{Brandt1999}.
However, such a scaling makes resulting $\sigma$ estimate independent of the ``scale'' of input variances $s_i^2$, and dependent only on their ratio.

So, both approaches give the same estimate for the WA, but different estimates for the WA uncertainty.
The first value $\sigma_1$ depends on $s_i$ and does not depend on the scatter of the input values $x_i$.
On the other hand, $\sigma_2$ depends on relative values of input variances $s_i^2$ and the scatter of $x_i$.
Difference between the two $\sigma$ estimates may be attributed to systematic errors in $x_i$ or underestimated $s_i$.

Theoretically, choice between $\sigma_1$ and $\sigma_2$ depends on whether the scatter of $x_i$ is a result of random error or
there are systematic differences between estimates $x_i$.
Obviously, both effects are present in most of practical applications.
This is a well recognized problem in data analysis, and its rigorous solution is hardly possible due to generally biased input estimates,
not always adequate reported uncertainties, and unknown error distribution.

In practice, if $x_i$ are close each other and $s_i$ are greater than the scatter of $x_i$, it seems reasonable to use $\sigma_1$.
Otherwise, if $s_i$ are much less than the scatter of the input measurements, $\sigma_2$ estimate seems to be more adequate to the data.
Indeed such a way to choose the best estimate for the WA uncertainty cannot be considered satisfactory.

A possible practical, statistically based approach has been proposed by Rosenfeld et al. (1967) and \citet{Brandt1999}.
According to this approach, $\chi^2$ criteria is used to decide whether the scatter of $x_i$ is a result of random errors.
First, both uncertainty estimates $\sigma_1$ and $\sigma_2$ are computed.
Then the final WA uncertainty is taken as
\begin{equation}
\sigma_3 = \left\{
\begin{array}{ll}
\sigma_1, & \ \mbox{if $H \le \chi^2(Q,n-1)$} \ , \\[1ex]
\sigma_2, & \ \mbox{if $H > \chi^2(Q,n-1)$} \ ,
\end{array}
\right.
\label{eq:sigma_3}
\end{equation}
where $Q$ is a significance level.
One can see that to a first approximation $\sigma_3 = \sigma_1$ for consistent measurements and $\sigma_3 = \sigma_2$ for discrepant ones,
and given $Q$ value is used to distinguish between them.
As a consequence, substantially different $\sigma$ estimates can be obtained for the same input measurements $(x_i, s_i)$ but specifying different $Q$.

Similar approach is discussed by \citet{Bich2002}.
It recommends accept $\sigma_1$ as the estimate of the CM mean uncertainty if consistency check $H \le \chi^2(Q,n-1)$ passed at a significance level
$Q$=5\%; otherwise supplement studies should be performed, such check of outliers, investigation of input data, etc.
Unfortunately, the latter is generally not feasible.

\subsection{Combined WA uncertainty estimator}

As pointed out in the previous section, the literature recommends using either $\sigma_1$ or $\sigma_2$ depending on some criteria, which can lead to ambiguous results,
keeping in mind that these two estimates may differ by several times.
So, a more robust $\sigma$ estimate is desirable for practical use, which would account of both the scatter of $x_i$ and their uncertainties $s_i$.
After investigation of behavior of all three estimates using simulated and real data, and supplement tests we decided in favor of combined estimate
computed by simple formula
\begin{equation}
\sigma_c = \sqrt{\sigma_1^2+\sigma_2^2} \ .
\label{eq:sigma_c}
\end{equation}

As a variant, $\sigma_2$ computed with unit weights can be used, which provides clear separation of impacts at combined estimate $\sigma_c$
from the uncertainties ($\sigma_1$) and scatter ($\sigma_2$) of the input measurements.
However, in this case, the result is generally more sensitive to measurements suspected to be outliers.

Unfortunately, we cannot suggest a rigorous theoretical ground of this approach, which is common for other practical recommendations too.
Our considerations are as follows.
Suppose we can represent each input value as $x_i=x+\varepsilon_{i}+\varepsilon_{0i}$, where $x$ is the true value of the CM, $\varepsilon_{i}$
is a random error distributed as $N(0,s_i^2)$ and $\varepsilon_{0i}$ is a systematic error of the $x_i$ measurement distributed as $N(0,\sigma_0^2)$.
Here $\sigma_0$ is considered as a measure of the scatter of the set of systematic errors in input measurements.
Evidently, $\varepsilon_{0i}$ is unknown, otherwise it would be accounted for in the reported value of $x_i$.
We can suppose that $\varepsilon_{0i}$ biases $x_i$ but does not bias $s_i$.
Thus the mathematical expectation of each $x_i$ is
\begin{equation}
{\cal{E}}(x_i)=x+\varepsilon_{0i} \, .
\label{eq:xi_expectation}
\end{equation}
Now we can use the set of $n$ equations~(\ref{eq:xi_expectation}) for $i=1,\ldots,n$ as an equations of condition to be solved by the least squares method.
As a result of this solution, we obtain an estimate of $x$ and its uncertainty $\sigma_0$, which can be expected close to $\sigma_2$.
Then we can consider $\sigma_0\approx\sigma_2$ as an additive error in the WA uncertainty.
Combining this error with $\sigma_1$ computed under the assumption of absence of systematic errors in $x_i$ we get $\sigma_c$ as defined by Eq~\ref{eq:sigma_c}.

Finally, let us notice that (\ref{eq:sigma_c}) can be rewritten as
\begin{equation}
\sigma_c = \sqrt{\fracd{1}{p}\left(1+\fracd{H}{n-1}\right)} \ .
\label{eq:sigma_c2}
\end{equation}
So, to obtain $\sigma_c$ estimate, there is no need to compute separately both $\sigma_1$ and $\sigma_2$ and then use Eq.~(\ref{eq:sigma_c}).

\subsection{Median}

Another approach routinely used to get the estimate of a CM is computation of a median $\bar{x}_m$.
The median is known as a robust statistics less influenced by outliers.
However its standard definition does not provide an estimate of error of a median value (it makes it immune to unreliable uncertainties though).

A possible approach to compute a median uncertainty was proposed by \citet{Muller1995,Muller2000a}.
Let $\bar{x}_{m}$ be the median of $x_i$, i.e. $\bar{x}_{m} = med\{x_i\}$.
Now we can compute the median of the absolute deviations (MAD) as
\begin{equation}
MAD = med\{|x_i - \bar{x}_{m}|\} \ .
\label{eq:mad}
\end{equation}

The uncertainty of $\bar{m}$ is then taken as
\begin{equation}
\sigma_m = \fracd{1.8582}{\sqrt{n-1}} \; MAD \ .
\label{eq:sigma_m}
\end{equation}

One can see that this estimate of the median uncertainty depends only on the data scatter and not on input uncertainties.
Later \citet{Muller2000b} proposed a method to take account of the uncertainty in input data and thus compute weighted median and its uncertainty.
However its practical realization, as pointed out by the author, is more cumbersome, and the testing results and discussion given therein
do not show clear advantage of using weighted median.

\section{Tests with artificial data}

In this section, results of two tests with simulated data are presented.
These tests were constructed to investigate in more details the behavior of the $\sigma$ estimates introduced in the previous section.
Indeed, many of features discussed here can be seen directly from corresponding equations, but not so demonstrative.

Table~\ref{tab:stat_examples} shows some numerical examples of computation of WA for two measurements, and its standard deviation.
To compute $\sigma_3$ we used $Q$=99\%, which corresponds to $\chi^2$(0.99,1)=6.635.
Using $Q$=95\%, $\chi^2$(0.95,1)=3.841 does not change main conclusions.
Note that, unlike general practice, we keep several significant digits in uncertainty just to better show the difference between various estimates.

Classical example 1 shows that $\sigma_2$ cannot provide reasonable estimate for $\sigma$ whatever how large input uncertainty are given.
Examples 2--7 and 8--15 show how $\sigma$ estimates change with grown $s_i$ for the same $x_i$.
Examples 16--23 show how $\sigma$ estimates change with grown $x_2$ for the same $x_1$ and $s_i$.
One can see that no one of $\sigma_1$, $\sigma_2$, $\sigma_3$ provides a satisfactory estimate of $\sigma$ for all the examples.

Several observations from Table~\ref{tab:stat_examples} are as follows.
The $\sigma_1$ estimate sometimes it is clearly underestimated (examples 8--12).
Examples 2--7 and 8--15 illustrate that $\sigma_2$ (2) cannot provide satisfactory estimate, especially in the cases 7, 14, 15, where it seems
to be underestimated.
Estimate $\sigma_3$ gives more realistic result, but not in all the cases, e.g. 2--3 and 8--11.
Moreover, $\sigma_3$ value depends not only on data sample $\{x_i,\, s_i\}$ but also on subjective choice of $Q$.
Besides, as can be seen from this test and Eq.~\ref{eq:sigma_3} that $\sigma_3$ may show significant jumps caused by small
changes in input data or confidence level.
For these reasons, it was decided not to use $\sigma_3$ in further work.

In contrast to $\sigma_1$, $\sigma_2$, and $\sigma_3$, one can see that $\sigma_c$ approach can provide stable and realistic estimate of
the standard deviation of the WA.

\begin{table*}
\centering
\caption{Examples of computation of WA of two measurements $x_1$, $x_2$ with associated uncertainties $s_1$, $s_2$.
Results of computations are mean $\bar{x}$ and four estimates of its uncertainty $\sigma_1$, $\sigma_2$, $\sigma_3$, and $\sigma_c$
computed by (\ref{eq:sigma_1}), (\ref{eq:sigma_2}), (\ref{eq:sigma_3}), and (\ref{eq:sigma_c}) respectively.
$H$ is computed by (\ref{eq:h}) and used to compute $\sigma_3$}
\label{tab:stat_examples}
\medskip
\begin{tabular}{|r|rr|r|r|r|rrrr|}
\hline
No. & $x_1$ & $x_2$ & $s_1, s_2$ & \multicolumn{1}{c|}{$\bar{x}$} & \multicolumn{1}{c|}{$H$} &
\multicolumn{1}{c}{$\sigma_1$} & \multicolumn{1}{c}{$\sigma_2$} &
\multicolumn{1}{c}{$\sigma_3$} & \multicolumn{1}{c|}{$\sigma_c$} \\
\hline
&&&&&&&&& \\
 1 &  1.0 & 1.0  &  0.5 &  1.0 &    0.00 &     0.354 & 0.000 & 0.354 & 0.354 \\
&&&&&&&&& \\
 2 &  1.0 &  2.0 &  0.1 &  1.5 &   50.00 &     0.071 & 0.500 & 0.500 & 0.505 \\
 3 &      &      &  0.2 &      &   12.50 &     0.141 & 0.500 & 0.500 & 0.520 \\
 4 &      &      &  0.3 &      &    5.56 &     0.212 & 0.500 & 0.212 & 0.543 \\
 5 &      &      &  0.5 &      &    2.00 &     0.354 & 0.500 & 0.354 & 0.612 \\
 6 &      &      &  1.0 &      &    0.50 &     0.707 & 0.500 & 0.707 & 0.866 \\
 7 &      &      &  2.0 &      &    0.12 &     1.414 & 0.500 & 1.414 & 1.500 \\
&&&&&&&&& \\
 8 & 10.0 & 20.0 &  0.1 & 15.0 & 5000.00 &     0.071 & 5.000 & 5.000 & 5.000 \\
 9 &      &      &  0.5 &      &  200.00 &     0.354 & 5.000 & 5.000 & 5.012 \\
10 &      &      &  1.0 &      &   50.00 &     0.707 & 5.000 & 5.000 & 5.050 \\
11 &      &      &  2.0 &      &   12.50 &     1.414 & 5.000 & 5.000 & 5.196 \\
12 &      &      &  3.0 &      &    5.56 &     2.121 & 5.000 & 2.121 & 5.431 \\
13 &      &      &  5.0 &      &    2.00 &     3.536 & 5.000 & 3.536 & 6.124 \\
14 &      &      & 10.0 &      &    0.50 &     7.071 & 5.000 & 7.071 & 8.660 \\
15 &      &      & 20.0 &      &    0.12 &    14.142 & 5.000 &14.142 &15.000 \\
&&&&&&&&& \\
16 & 10.0 & 10.0 & 1.0  & 10.0 &   0.00  &     0.707 & 0.000 & 0.707 & 0.707 \\
17 & 10.0 & 11.0 &      & 10.5 &   0.50  &     0.707 & 0.500 & 0.707 & 0.866 \\
18 & 10.0 & 12.0 &      & 11.0 &   2.00  &     0.707 & 1.000 & 0.707 & 1.225 \\
19 & 10.0 & 13.0 &      & 11.5 &   4.50  &     0.707 & 1.500 & 0.707 & 1.658 \\
20 & 10.0 & 14.0 &      & 12.0 &   8.00  &     0.707 & 2.000 & 2.000 & 2.121 \\
21 & 10.0 & 15.0 &      & 12.5 &  12.50  &     0.707 & 2.500 & 2.500 & 2.598 \\
22 & 10.0 & 16.0 &      & 13.0 &  18.00  &     0.707 & 3.000 & 3.000 & 3.082 \\
23 & 10.0 & 17.0 &      & 13.5 &  24.50  &     0.707 & 3.500 & 3.500 & 3.571 \\
&&&&&&&&& \\
\hline
\end{tabular}
\end{table*}

\begin{figure*}
\centering
\begin{tabular}{c@{\hspace{5em}}l@{\hspace{5em}}c}
{\large a} &
\begin{tabular}{l}
$x_1 = 20.0 \pm 1.4$ \\
$x_2 = 18.7 \pm 1.7$ \\
$x_3 = 29.8 \pm 1.4$ \\
$x_4 = 21.1 \pm 1.6$ \\
$x_5 = 23.3 \pm 1.0$ \\[1ex]
\hline \\[-2ex]
$\bar{x}_{w1} = 23.00 \pm 0.60$ \\
$\bar{x}_{w2} = 23.00 \pm 1.81$ \\
$\bar{x}_{wc} = 23.00 \pm 1.91$ \\
$\bar{x}_{m} = 21.10 \pm 2.04$
\end{tabular} &
\begin{tabular}{c}
\\[1ex]
\includegraphics[clip,width=0.5\hsize]{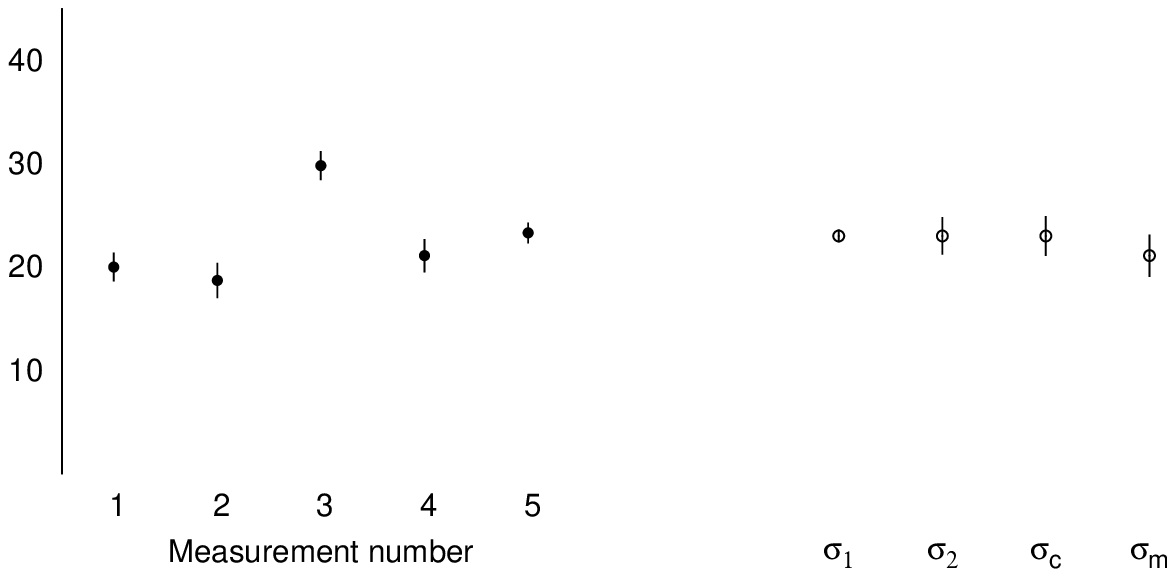}
\end{tabular} \\
&& \\[1em]
{\large b} &
\begin{tabular}{l}
$x_1 = 20.0 \pm 4.2$ \\
$x_2 = 18.7 \pm 5.1$ \\
$x_3 = 29.8 \pm 4.2$ \\
$x_4 = 21.1 \pm 4.8$ \\
$x_5 = 23.3 \pm 3.0$ \\[1ex]
\hline \\[-2ex]
$\bar{x}_{w1} = 23.00 \pm 1.81$ \\
$\bar{x}_{w2} = 23.00 \pm 1.81$ \\
$\bar{x}_{wc} = 23.00 \pm 2.56$ \\
$\bar{x}_{m} = 21.10 \pm 2.04$
\end{tabular} &
\begin{tabular}{c}
\\[1ex]
\includegraphics[clip,width=0.5\hsize]{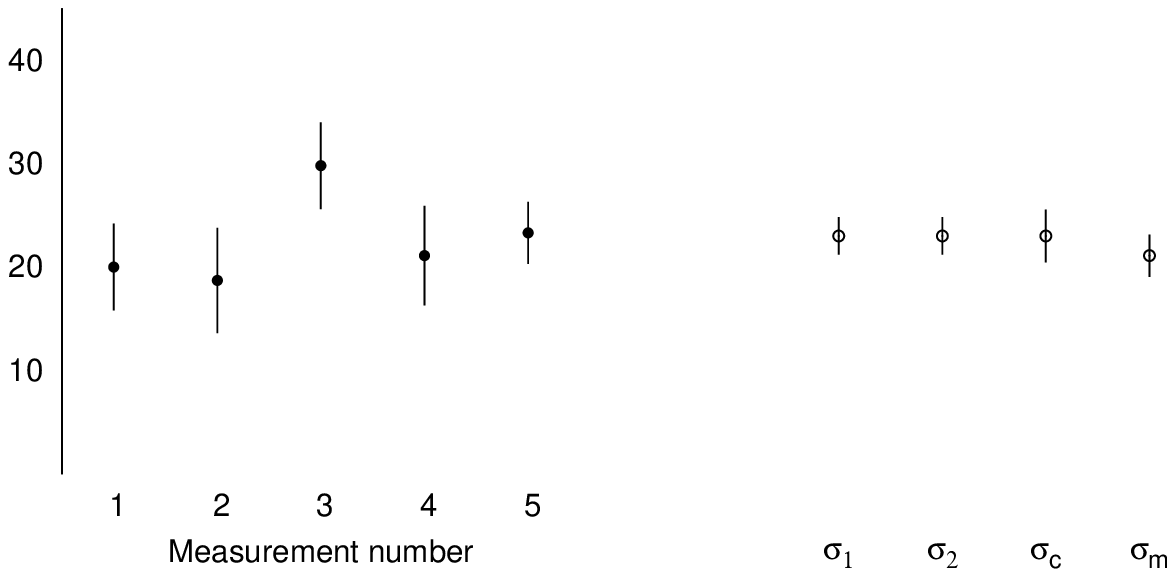}
\end{tabular} \\
&& \\[1em]
{\large c} &
\begin{tabular}{l}
$x_1 = 20.0 \pm 12.6$ \\
$x_2 = 18.7 \pm 15.3$ \\
$x_3 = 29.8 \pm 12.6$ \\
$x_4 = 21.1 \pm 14.4$ \\
$x_5 = 23.3 \pm ~9.0$ \\[1ex]
\hline \\[-2ex]
$\bar{x}_{w1} = 13.00 \pm 5.42$ \\
$\bar{x}_{w2} = 13.00 \pm 1.81$ \\
$\bar{x}_{wc} = 13.00 \pm 5.71$ \\
$\bar{x}_{m} = 11.10 \pm 2.04$
\end{tabular} &
\begin{tabular}{c}
\\[1ex]
\includegraphics[clip,width=0.5\hsize]{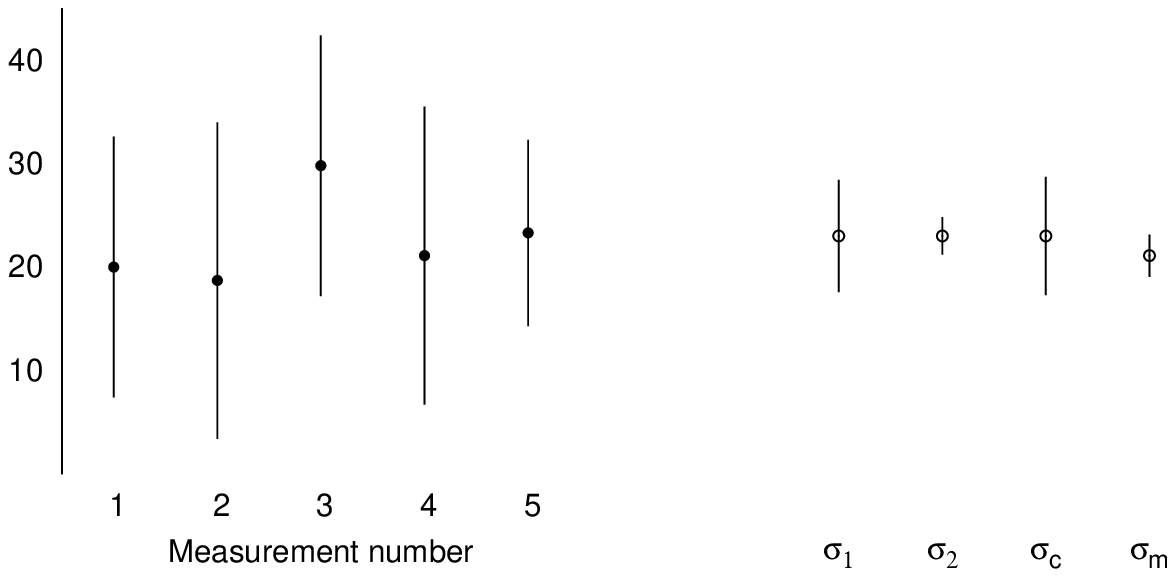}
\end{tabular} \\
\end{tabular}
\caption{Testing four methods to compute CM with simulated data.
The left group of points (discs) represents measurements, and the right group of points (circles) represents CM estimates.
In all three examples, averaged measurements $x_i$ are the same, but their uncertainties differ.
Uncertainties in example $b$ are three times greater than those in example $a$, and uncertainties in example $c$ are three times greater than those in example $b$;
$\bar{x}_{w1}$ is the WA computed by Eq.~(\ref{eq:basic}) with the uncertainty $\sigma_1$ computed by (\ref{eq:sigma_1}).
$\bar{x}_{w2}$ is the same WA with the uncertainty $\sigma_2$ computed by (\ref{eq:sigma_2}),
$\bar{x}_{wc}$ is the same WA with the uncertainty $\sigma_c$ computed by (\ref{eq:sigma_c}),
$\bar{x}_{m}$ is the median with the uncertainty $\sigma_m$ computed by (\ref{eq:sigma_m})
}
\label{fig:simulation}
\end{figure*}

In the second test, we use the same set of five measurements with different errors (see Fig.~\ref{fig:simulation}).
In the upper row the data have minimal uncertainties; in the middle row, all the uncertainties are increased by factor of 3;
and in the bottom row, all the uncertainties are increased again by factor of 3.

From this Figure, one can see that $\sigma_1$ became greater as $s_i$ grow, as expected.
However it looks underestimated in case $a$.
The $\sigma_2$ estimate remains the same for all three examples because it does not depend on absolute values of $s_i$, but only on the dispersion ratio,
which is the same in all the cases.
One would expect however that $\sigma$ should be greater in two last examples as compared to the previous ones.
Median estimate is close to $\sigma_2$ as expected because both of them depend on the $x_i$ scatter only, and hence $\sigma_m$ shows the same problems as $\sigma_2$.

The $\sigma_c$ estimate appears to be optimal because it shows a steady increase from case $a$ to case $c$ accounting both for input $s_i$ value and $x_i$ scatter.
In case $a$ with small $s_i$ values $\sigma_c$ is determined mainly by the data scatter, and is close to $\sigma_2$.
In case $b$ with equal contribution of input data uncertainties and scatter ($\sigma_1=\sigma_2$), $\sigma_c$ is just greater by factor of $\sqrt{2}$.
In case $c$, $\sigma_c$ is defined mainly by $s_i$, which are much greater than the data scatter, and it is close to $\sigma_1$.
We can say that the $\sigma_c$ estimate ``automatically'' takes account of both input measurements scatter and uncertainties without any need in
supplement assumptions or parameters like significance level.

\section{Application to real data}

\begin{figure*}
\centering
\begin{tabular}{l@{\hspace{5em}}c}
\begin{tabular}{l}
$x_1 = 3847.5 \pm 0.4$ \\
$x_2 = 3847.9 \pm 0.4$ \\
$x_3 = 3846.6 \pm 0.5$ \\
$x_4 = 3848.1 \pm 0.2$ \\[1ex]
\hline \\[-2ex]
$\bar{x}_{w1} = 3847.83 \pm 0.16$ \\
$\bar{x}_{w2} = 3847.83 \pm 0.26$ \\
$\bar{x}_{wc} = 3847.83 \pm 0.31$ \\
$\bar{x}_{m} = 3847.70 \pm 0.32$
\end{tabular} &
\begin{tabular}{c}
\\[1ex]
\includegraphics[clip,width=0.5\hsize]{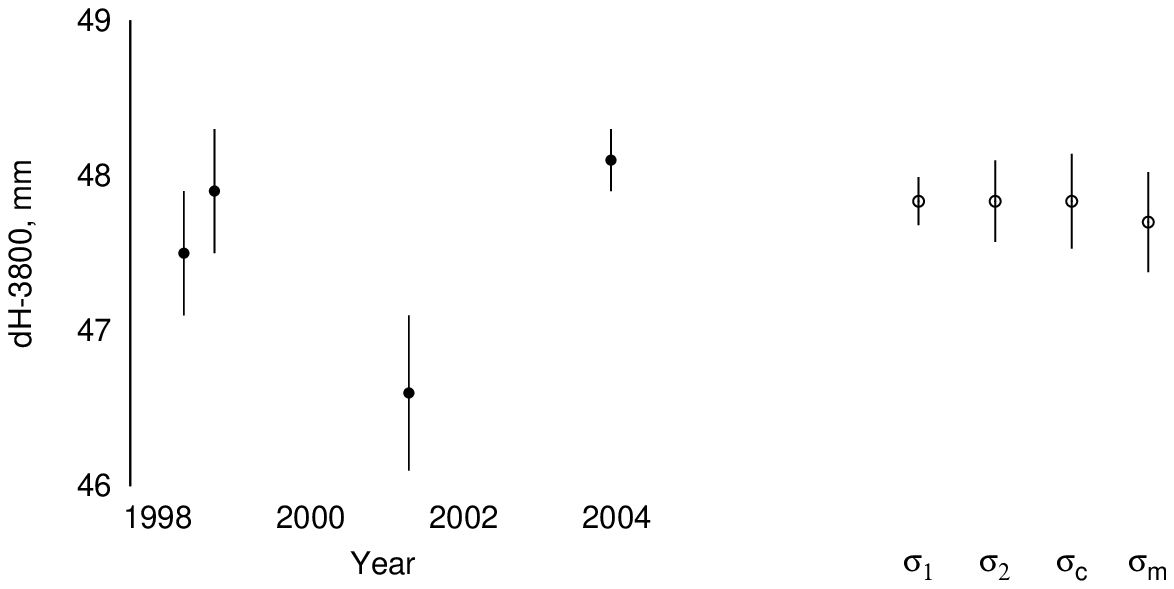}
\end{tabular} \\
\end{tabular}
\caption{Testing four methods to compute a CA with real data: a case of determination of the height difference between two geodetic marks.
The designations are the same as in Fig~\ref{fig:simulation}}
\label{fig:Sv_dh}
\end{figure*}

In this section, the tests with real data are presented.
In the first test, the height differences are analyzed between marks 107 and 109 of the local geodetic network of the Svetloe Radio Astronomical Observatory
of the Institute of Applied Astronomy, St.~Petersburg, Russia \citep{Kazarinov1997,Finkelstein2006}.
This analysis includes four levelling surveys performed in 1998--2003.
Distance between the marks is about 135~m.
Results of computations are shown in Fig~\ref{fig:Sv_dh}.
In this case, the scatter of the measurements is rather large as compared to measurement uncertainties.
For this reason, $\sigma_1$ seems to be underestimated.
Combined estimate $\sigma_3$ is close to $\sigma_2$, but may be preferable because accounts also for input uncertainties.
Median uncertainty is close to $\sigma_2$.
In this example, it is difficult to decide which of the two latter estimates should be preferred.
Both of them looks equally realistic.

\begin{figure*}
\centering
\begin{tabular}{l@{\hspace{5em}}c}
\begin{tabular}{l}
$x_1 = 15.0 \pm 0.8$ \\
$x_2 = 14.4 \pm 1.2$ \\
$x_3 = 11.3 \pm 1.1$ \\
$x_4 = 14.8 \pm 0.8$ \\
$x_4 = 14.5 \pm 1.5$ \\[1ex]
\hline \\[-2ex]
$\bar{x}_{w1} = 14.21 \pm 0.44$ \\
$\bar{x}_{w2} = 14.21 \pm 0.65$ \\
$\bar{x}_{wc} = 14.21 \pm 0.79$ \\
$\bar{x}_{m} = 14.50 \pm 0.28$
\end{tabular} &
\begin{tabular}{c}
\\[1ex]
\includegraphics[clip,width=0.5\hsize]{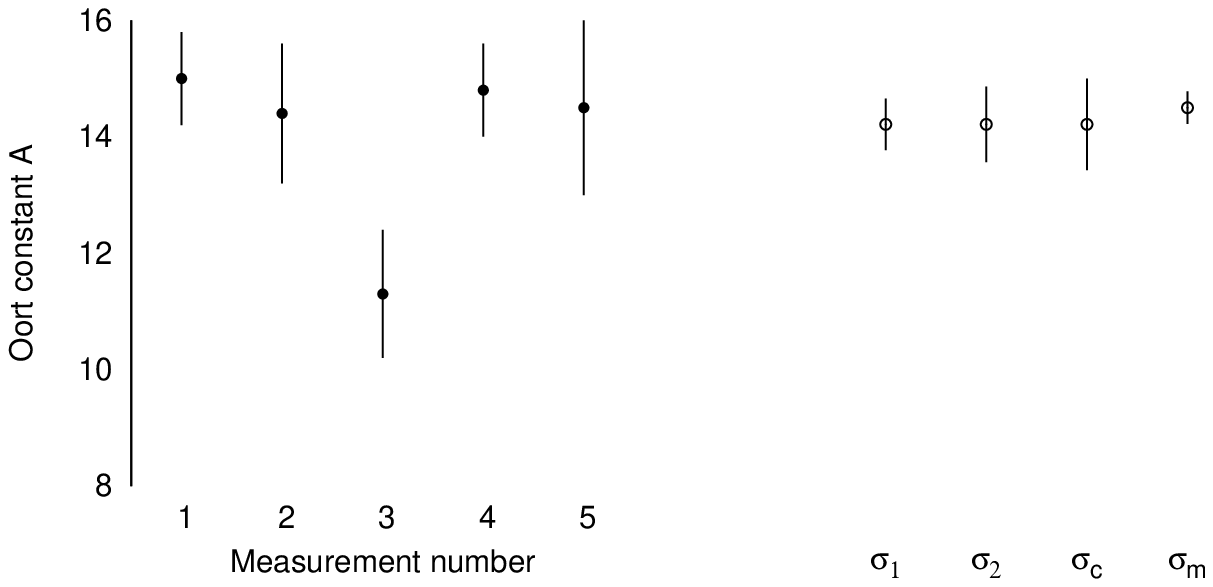}%
\end{tabular} \\
&\\
\begin{tabular}{l}
$x_1 = -10.0 \pm 1.2$ \\
$x_2 = -12.0 \pm 2.8$ \\
$x_3 = -13.9 \pm 0.9$ \\
$x_4 = -12.4 \pm 0.6$ \\
$x_4 = -12.0 \pm 3.0$ \\[1ex]
\hline \\[-2ex]
$\bar{x}_{w1} = -12.42 \pm 0.45$ \\
$\bar{x}_{w2} = -12.42 \pm 0.59$ \\
$\bar{x}_{wc} = -12.42 \pm 0.74$ \\
$\bar{x}_{m} = -12.00 \pm 0.37$
\end{tabular} &
\begin{tabular}{c}
\\[1ex]
\includegraphics[clip,width=0.5\hsize]{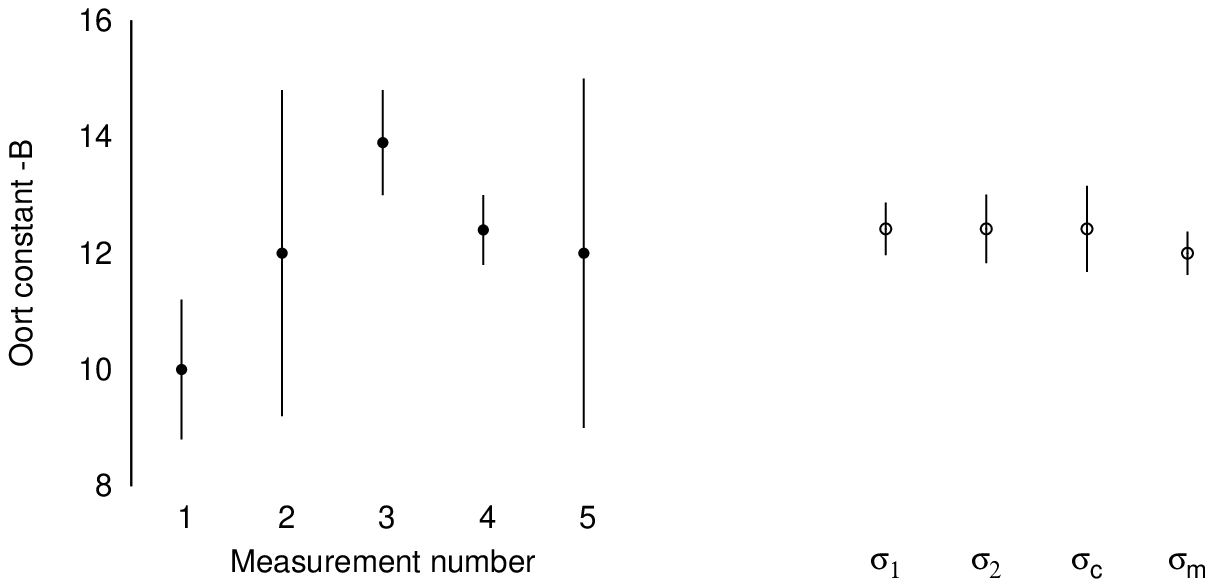}%
\end{tabular} \\
\end{tabular}
\caption{Testing four methods to compute a CA with real data: a case of determination of the Oort constants A (top) and B (bottom).
The designations are the same as in Fig~\ref{fig:simulation}}
\label{fig:Oort}
\end{figure*}

In the second test, results of determination of average values of the Oort constants in \citet{Klacka2009} are revised
(see Fig~\ref{fig:Oort}).
For comparison, the author's estimates are $14.2 \pm 0.5$ for A, and $-12.4 \pm 0.5$ for B,
with uncertainties computed as $\sigma_1$ and evidently rounded up.
This value of the WA uncertainty is likely underestimated as compared to input data uncertainties and scatter.
Median uncertainty also seems underestimated because it accounts only for relatively small scatter, and ignores relatively large measurement
uncertainties.
In this example, $\sigma_c$ estimate again looks the most realistic and corresponding to input data.

\section{Conclusion}

Although computation of a CM, in particular WA, is widely used in data analysis, this problem has no definite, unambiguity solution yet.
In particular, a very important problem in most of applications is to obtain a ``realistic'' estimate for the CM uncertainty.
Both underestimation and overestimation are equally undesirable.

In this study, we mostly investigated several basic approaches to compute a WA.
Currently used methods for computation of the WA uncertainty do not provide satisfactory result for many practical tasks.
The classical WA uncertainty estimate $\sigma_1$ often yields underestimated value because it is rigorously justified for unbiased $x_i$ only.
Another estimate $\sigma_2$, which can be derived from a least squares solution or by appropriate scaling of the input uncertainties, does not take
account of input uncertainties, but only of the variance ratio.

In this paper, we propose a new approach to compute the WA uncertainty, combined estimate $\sigma_c$ given by (\ref{eq:sigma_c}) and (\ref{eq:sigma_c2}),
which is able to account for both uncertainties and scatter of input data.
We did not propose a rigorous statistical background for this estimate.
However, it can be shown that in the case when input values $x_i$ are obtained from a normally distributed populations, and each value $x_i$ has
a normally distributed systematic error, proposed estimate $\sigma_c$ can be derived from a least squares solution.
It appears probable that this estimate is suitable for practical use until the deviation from normality is very large.

Several tests with simulated and real measurements have demonstrated that using $\sigma_c$ is a simple and effective approach equally suitable for both
discrepant and consistent measurements.
It is also important that it provides realistic $\sigma$ estimate even for very small samples of 2--3 measurements.

As to the median approach, it is known as a more robust CM estimate, but computation of its uncertainty is not so straightforward.
A simple approach by \citet{Muller1995,Muller2000a} seems to be not satisfactory in some cases, as follows from our tests.
In this respect, the bootstrap method and extended bootstrap method \citep{MacMahon2004} deserve consideration; this methods may be too complicated
for routine use though.

In conclusion, one should never forget the following.
Despite the method used, the CM uncertainty is only a part of real measurement accuracy, namely a Type A uncertainty according to the standard
metrological terminology.
By definition, Type A uncertainty is computed from data using statistical procedures, while a Type B uncertainty is obtained using supplement information
and theoretical considerations (see e.g. \citet{Bucher2004}.
Sometimes, the type B uncertainty is evaluated from supplement testing data processing, e.g. using subsets of the original data set.
It can also be a result not of mathematical computation, but considerations based on the knowledge of the measurement procedures, observational history,
and previous experience.
Such procedures were used e.g. to derive the best estimates of some quantities related to Solar System dynamics \citep{Pitjeva2009},
and to compute realistic errors in radio source positions for the second realization of the International Celestial Reference Frame \citep{Ma2009p}.

\bibliography{}

\begin{thebibliography}{27}
\providecommand{\natexlab}[1]{#1}
\providecommand{\doi}[1]{doi:\discretionary{}{}{}#1}
\providecommand{\url}[1]{{#1}}
\providecommand{\eprint}[2][]{\url{#2}}

\bibitem[{{Bevington} and {Robinson}(2003)}]{Bevington2003}
{Bevington} PR, {Robinson} DK (2003) Data reduction and error analysis for the
  physical sciences, 3rd ed. McGraw-Hill, Boston

\bibitem[{{Bich} et~al(2002){Bich}, {Cox}, {Estler}, {Nielsen}, and
  {Woeger}}]{Bich2002}
{Bich} W, {Cox} M, {Estler} T, {Nielsen} L, {Woeger} W (2002) Draft for
  discussion. Proposed guidelines for the evaluation of key comparison data.
  \url{http://www.bipm.org/cc/CCAUV/Allowed/3/CCAUV02-36.pdf}

\bibitem[{{Brandt}(1999)}]{Brandt1999}
{Brandt} S (1999) Data analysis: statistical and computational methods for
  scientists and engineers, 3rd ed. Springer

\bibitem[{{Bucher}(2004)}]{Bucher2004}
{Bucher} JL {(ed)} (2004) The Metrology Handbook. ASQ Quality Press, Milwaukee,
  Wisconsin

\bibitem[{{Chen} et~al(2011){Chen}, {Geraedts}, {Ouellet}, and
  {Singh}}]{Chen2011}
{Chen} J, {Geraedts} SD, {Ouellet} C, {Singh} B (2011) Evaluation of half-life
  of $^{198}$Au. Appl Rad Isot 69:1064--1069.
  \doi{10.1016/j.apradiso.2011.03.024}

\bibitem[{{Cox} and {Harris}(2004)}]{Cox2004}
{Cox} MG, {Harris} PM (2004) Technical aspects of guidelines for the evaluation
  of key comparison data. Meas Tech 47:102--111.
  \doi{10.1023/B:METE.0000022513.17564.5b}

\bibitem[{{Figueiredo} and {Gomes}(2004)}]{Figueiredo2004}
{Figueiredo} F, {Gomes} M (2004) The total median in statistical quality
  control. Appl Stoch Models Bus Ind 20:339--353. \doi{10.1002/asmb.545}

\bibitem[{{Finkelstein} et~al(2006){Finkelstein}, {Ipatov}, {Malkin},
  {Skurikhina}, and {Smolentsev}}]{Finkelstein2006}
{Finkelstein} AM, {Ipatov} AV, {Malkin} ZM, {Skurikhina} EA, {Smolentsev} S
  (2006) Results of the first two years of VLBI observations at Svetloe
  observatory within in the framework of international geodynamical programs.
  Astron Lett 32:138--144. \doi{10.1134/S1063773706020083}

\bibitem[{{Graybill} and {Deal}(1959)}]{Graybill1959}
{Graybill} FA, {Deal} RB (1959) Combining unbiased estimators. Biometrics
  15:543--550

\bibitem[{{Helene}(2007)}]{Helene2007}
{Helene} O (2007) Comparison of the bootstrap method with another method for
  the analysis of discrepant data sets. Nucl Instrum Methods A 574:144--149.
  \doi{10.1016/j.nima.2007.01.183}

\bibitem[{{Kazarinov} and {Malkin}(1997)}]{Kazarinov1997}
{Kazarinov} AS, {Malkin} ZM (1997) Geodetic service of the VLBI network QUASAR.
  Transactions of IAA RAS 2:286--299

\bibitem[{{Kla\v{c}ka}(2009)}]{Klacka2009}
{Kla\v{c}ka} J (2009) {Galactic tide}. arXiv 0912.3112

\bibitem[{{MacMahon} et~al(2004){MacMahon}, {Pearce}, and
  {Harris}}]{MacMahon2004}
{MacMahon} D, {Pearce} A, {Harris} P (2004) Convergence of techniques for the
  evaluation of discrepant data. Appl Rad Isot 60:275--281.
  \doi{10.1016/j.apradiso.2003.11.028}

\bibitem[{{Malkin}(2011)}]{Malkin2011fe}
{Malkin} ZM (2011) The influence of Galactic aberration on precession
  parameters determined from VLBI observations. Astron Rep 55:810--815.
  \doi{10.1134/S1063772911090058}

\bibitem[{{Malkin}(2001{\natexlab{a}})}]{Malkin2001m}
{Malkin} Z (2001{\natexlab{a}}) On computation of combined IVS EOP series. In:
  {Behrend} D, {Rius} A (eds) 15th Workshop Meeting on European VLBI for
  Geodesy and Astrometry, pp 55--62

\bibitem[{{Malkin}(2001{\natexlab{b}})}]{Malkin2001k}
{Malkin} Z (2001{\natexlab{b}}) On computation of weighted mean. Communications
  of the Institute of Applied Astronomy RAS, No~137

\bibitem[{{Malkin}(2008)}]{Malkin2008g}
{Malkin} Z (2008) On construction of ICRF-2. In: {Finkelstein} A, {Behrend} D
  (eds) Measuring the Future, Proceedings of the Fifth IVS General Meeting, St.
  Petersburg, Russia, March 2-6, 2008. St. Petersburg, 2008. ISBN
  978-5-02-025332-2, pp 256--260

\bibitem[{{Ma} et~al(2009){Ma}, {Arias}, {Bianco}, {Boboltz}, {Bolotin},
  {Charlot}, {Engelhardt}, {Fey}, {Gaume}, {Gontier}, {Heinkelmann}, {Jacobs},
  {Kurdubov}, {Lambert}, {Malkin}, {Nothnagel}, {Petrov}, {Skurikhina},
  {Sokolova}, {Souchay}, {Sovers}, {Tesmer}, {Titov}, {Wang}, {Zharov},
  {Barache}, {Boeckmann}, {Collioud}, {Gipson}, {Gordon}, {Lytvyn},
  {MacMillan}, and {Ojha}}]{Ma2009p}
{Ma} C, {Arias} EF, {Bianco} G, {Boboltz} DA, {Bolotin} SL, {Charlot} P,
  {Engelhardt} G, {Fey} AL, {Gaume} RA, {Gontier} AM, {Heinkelmann} R, {Jacobs}
  CS, {Kurdubov} S, {Lambert} SB, {Malkin} ZM, {Nothnagel} A, {Petrov} L,
  {Skurikhina} E, {Sokolova} JR, {Souchay} J, {Sovers} OJ, {Tesmer} V, {Titov}
  OA, {Wang} G, {Zharov} VE, {Barache} C, {Boeckmann} S, {Collioud} A, {Gipson}
  JM, {Gordon} D, {Lytvyn} SO, {MacMillan} DS, {Ojha} R (2009) The second
  realization of the International Celestial Reference Frame by Very Long
  Baseline Interferometry. In: {Fey} AL, {Gordon} D, {Jacobs} CS (eds) IERS
  Technical Note No~35, Verlag des Bundesamts fuer Kartographie und Geod\"asie,
  Frankfurt am Main

\bibitem[{{M\"uller}(1995)}]{Muller1995}
{M\"uller} JW (1995) Possible advantages of a robust evaluation of comparisons.
  Report BIPM-95/2. Bureau International des Poids et Mesures, S\`evres, France

\bibitem[{{M\"uller}(2000{\natexlab{a}})}]{Muller2000a}
{M\"uller} JW (2000{\natexlab{a}}) Possible advantages of a robust evaluation
  of Comparisons. J Res Natl Inst Stand Technol 105:551--555.
  \doi{10.1016/j.apradiso.2003.11.028}

\bibitem[{{M\"uller}(2000{\natexlab{b}})}]{Muller2000b}
{M\"uller} JW (2000{\natexlab{b}}) Weighted median. Report BIPM-2000/6. Bureau
  International des Poids et Mesures, S\`evres, France

\bibitem[{Pitjeva~EV(2009)}]{Pitjeva2009}
Pitjeva~EV SE (2009) Proposals for the masses of the three largest asteroids,
  the Moon-Earth mass ratio and the Astronomical Unit. Cel Mech Dyn Astr
  103:365--372. \doi{10.1007/s10569-009-9203-8}

\bibitem[{{Rosenfeld} et~al(1967){Rosenfeld}, {Barbaro-Galtieri}, {Podolsky},
  {Price}, {Soding}, {Wohl}, {Roos}, and {Willis}}]{Rosenfeld1967}
{Rosenfeld} AH, {Barbaro-Galtieri} A, {Podolsky} WJ, {Price} LR, {Soding} P,
  {Wohl} CG, {Roos} M, {Willis} WJ (1967) Data on particles and resonant
  states. Rev Mod Phys 39:1--51

\bibitem[{{Sinha}(1985)}]{Sinha1985}
{Sinha} B (1985) Unbiased estimation of the variance of the Graybill-Deal
  estimator of the common mean of several normal populations. Canadian J of
  Statistics 13:243--247. \doi{10.2307/3315154}

\bibitem[{{Sokolova} and {Malkin}(2007)}]{Sokolova2007}
{Sokolova} J, {Malkin} Z (2007) On comparison and combination of catalogues of
  radio source positions. Astron. Astrophys. 474:665--670. \doi{10.1051/0004-6361:20077450}

\bibitem[{{Witkovsk\'y} and {Wimmer}(2001)}]{Witkovsky2001}
{Witkovsk\'y} W, {Wimmer} G (2001) On statistical models for consensus values.
  Meas Sci Rev~1 (Section 1):33--36

\bibitem[{{Zhang}(2006)}]{Zhang2006}
{Zhang} N (2006) The uncertainty associated with the weighted mean of
  measurement data. Metrologia 43:195--204. \doi{10.1088/0026-1394/43/3/002}

\end{thebibliography}
\bibliographystyle{joge}

\end{document}